\documentclass{ws-procs975x65}

\usepackage{epsfig}

\begin{document}
\title{Size and charge effects on the deposition of Na on Ar}
\author{P.~M.~Dinh$^1$, F.~Fehrer$^2$, P.-G.~Reinhard$^2$, and
  E.~Suraud$^1$} 
\address {$^1$Laboratoire de Physique Th\'eorique,
        UMR 5152, Universit\'e Paul Sabatier,
        118 route de Narbonne, F-31062 Toulouse Cedex,
        France}
\address {$^2$Institut f\"ur Theoretische Physik, Universit\"at Erlangen
              Staudtstr.\ 7, D-91058 Erlangen, Germany}

\begin{abstract}
We discuss the dynamical deposition of the Na atom, the Na$^+$ in and 
the Na$-6$ cluster on finite Ar clusters mocking up an infinite Ar surface. 
We analyze this scenario as a function of projectile initial 
kinetic energy and of the size the target cluster.
\end{abstract}


\section{Introduction}

Clusters on surfaces have motivated many  studies over the 
past decades and still remain a topic of great interest \cite{Bru00},
especially in relation to the 
synthesis of nanostructured
surfaces~\cite{ISSPIC9,ISSPIC10,ISSPIC11,ISSPIC12}. Indeed, it turns out
that it is  
 possible to make a  direct deposition of size selected clusters on a
substrate \cite{Bin01,Har00}. The deposition process may lead to 
a significant modification of the cluster, in terms of its
electronic structure and ionic geometry. This is 
a consequence of the impact of the interface energy, the
electronic band structure of the substrate, and the surface corrugation. 
These questions have already been widely investigated in great detail, 
especially from the structural point of view and both from  the 
experimental \cite{Exp1,Exp2,Exp3} and theoretical 
\cite{BL,CL,HBL,MH,SurfPot1,SurfPot2,2harm,IJMS} sides.
The situation is somewhat different concerning the deposition process itself.
Its theoretical description remains mostly limited to 
molecular dynamics (MD) approaches, which implies that 
a proper description of electronic degrees of freedom is missing.
The reason for this defect is simple.
The presence of a substrate makes the experimental handling of clusters 
easier but strongly complicates the theoretical description because of the 
huge number of degrees of freedom of the surface. 
The MD then provides the cheapest way to access, at least in a gross way, 
the dynamics of the substrate. It is nevertheless crucial to try 
to account for the surface's electronic degrees of freedom, for example
when non adiabatic processes are involved.

As a first step in the direction of a fully microscopic 
dynamical approach, one may consider relatively 
simple cluster/substrate combinations. This is for example the case 
of the deposit of a metal cluster on an insulator surface. 
The surface can then be included at a
lower level of description, which simplifies the handling, as was, e.g.,
explored for the case of  Na clusters on NaCl in \cite{IJMS,Ipa04}. 
Cluster electrons were there described by means of Density Functional 
Theory  and the coupling to the surface was achieved {\it via}
an effective interface potential, itself tuned to {\it ab-initio}
calculations 
\cite{MH}.  Such an approach implies a total freezing of the 
surface itself, which sets severe limitations on its applicability. 
A somewhat better description
of surface degrees of freedom in a still limited/simplified way
can be achieved by considering again a rather 
inert substrate, but allowing for a minimum of dynamical response of
the substrate. 
Such a model was recently proposed for describing sodium clusters embedded in
rare gases \cite{Ger04b,Feh06a,Feh05c}. The method used was  "hierarchical", 
which is justified by the moderate interactions between
cluster and surface. The interface can then be treated at a simpler
level than the cluster's degrees of freedom
(microscopic treatment of cluster and classical treatment of environment with 
explicit account of dynamical polarizability effects), in the spirit of 
the coupled quantum-mechanical with
molecular-mechanical method (QM/MM) often used in
bio-chemistry \cite{Fie90a,Gao96a,Gre96a}. 

Such hierarchical methods, although much simpler than a fully
microscopic approach, require sophisticated modeling and are thus
restricted to finite systems. Nonetheless, the calculations in
\cite{Feh05c,Feh06a} were carried forth to a sufficiently large range
of sizes to see the appearance of generic behaviors on the way towards
the bulk.  In the complementing case of deposit on a surface, we also
consider finite substrates, as model cases for a surface. This is an
acceptable compromise once the impact of the finiteness of the
substrate has been properly analyzed.  For a first exploration, we
went one step further and restricted the present analysis to the even
simpler case of the deposition of a single Na atom on a finite Ar
cluster, adding one test case in which we shall explore the case of a
finite Na cluster.

The goal of this paper is thus mostly the study of the dynamics of deposition 
of a sodium atom (projectile) on Ar clusters of various 
sizes Ar$_N$(target, $N=43,86$). We shall analyze the behaviors of both 
the atom and the cluster, especially as a function of deposit velocity
and also  consider size and charge effects (deposit of a Na$_6$
cluster and a Na$^+$ ion instead of a Na atom).  
The paper is organized as follows. Section \ref{sec:model} gives a short 
presentation of the model used. A few more details on the model can be 
found in the appendix, where basic formulae and parameters are recalled. 
The following sections successively address the dependence on substrate 
size and on projectile velocity. We finally discuss the example of a true 
cluster deposit and of charge effects. 

\section{Model}
\label{sec:model}

The model was 
presented in detail in  \cite{Feh05a} and we 
provide the basic formulae in the appendix.
We recall briefly the basic ingredients. 
The Na cluster is described in terms of time-dependent
local-density approximation for the electrons coupled to molecular
dynamics for the ions (TDLDA-MD), a scheme which has been extensively
validated for linear and non-linear dynamics in free metal clusters
\cite{Rei03a,Cal00}.   
The electron-ion interaction
is treated by means of a soft, local pseudo-potentials \cite{Kue99}. 
Each Ar atom is described by two classical degrees-of-freedom: its
center-of-mass and its electrical dipole moment. The explicit account 
of Ar dipoles allows to 
treat the polarizability of the atoms dynamically, 
with help of  polarization potentials \cite{Dic58}. 
Atom-atom interactions are described by a standard Lennard-Jones potential.
For the Ar-Na$^+$ interaction we employ effective potentials from
the literature \cite{Rez95}. The
electron-Ar core repulsion is modeled in 
the form proposed by \cite{Dup96}, with a slight 
final readjustment of the parameters to the NaAr molecule as 
benchmark (bond length, binding
energy and optical excitation).
A Van der Waals interaction is also added and computed {\it via} the variance
of dipole operators \cite{Ger04b,Feh05a,Dup96}.

\begin{figure}[htbp]
\begin{center}
\epsfig{file=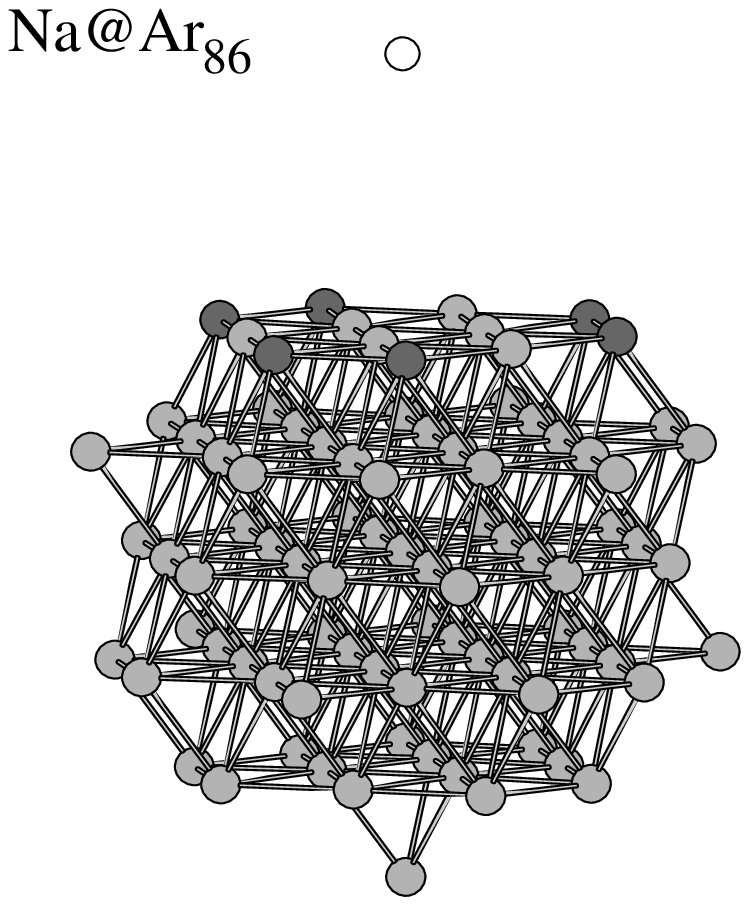,width=6cm}
\caption{\label{fig:config}
Atomic configuration of Ar$_{86}$ with respect to 
the single Na (white ball). Six Ar atoms are emphasized
by dark gray balls for the discussion on Na$^+$ deposition in
Ar$_{86}$ (see text).} 
\end{center}
\end{figure}

As explained in the appendix, 
the starting quantity is the total energy, 
constructed from the various pieces discussed above.
The corresponding equations of motion
can then be derived by standard variation, which 
leads to the time-dependent
Kohn-Sham equations for the cluster electrons. One furthermore
obtains  Hamiltonian equations of motion for the classical degrees 
of freedom (Na$^+$ ions as well as Ar
atom positions and dipoles). The initial condition
is  provided by the corresponding stationary solutions. 

The definition of the Ar-Na configuration
requires some specific handling. Indeed the Na atom can be initially
placed above either an Ar atom or an interstitial site of the
surface layer.
In the case of Ar$_{43}$ (see Figure~\ref{fig:NaAr43}), the first
option has been chosen, and for 
Ar$_{86}$, the second initial configuration has been used,
see Figure~\ref{fig:config}. 
The Ar$_{86}$ is obtained from the stable free Ar$_{87}$ cluster where an
outer Ar atom has been removed. 
The resulting Ar$_{86}$ is then rotated in order to present a flat
surface to the impinging Na atom, this way simulating the flat 
interface provided by an infinite surface.
The starting configuration for the deposit of Na$_6$ on Ar$_{43}$
is similar to that of the single Na atom. This means the top Na
ion (the Na$_6$ is composed of a ring of 5 ions and an outer ion) is
placed above an Ar atom.

The numerical solution proceeds with standard methods as 
detailed in \cite{Cal00}.
The Kohn-Sham equations for the cluster electrons are
solved using real space grid techniques. The time 
propagation proceeds using a time-splitting method. The stationary
solution is attained by accelerated gradient iterations. 
We furthermore employ the cylindrically-averaged pseudo-potential scheme
(CAPS) as introduced in \cite{Mon94a,Mon95a}, an approximation
justified for the chosen test cases. In the following, the symmetry
axis is denoted by the $z$ axis. It should nevertheless 
be noted that the dynamics of the Na$^+$ ions
as well as that of the Ar atoms are treated in full 3D.

\section{Dynamical deposition of Na on finite Ar clusters}

\subsection{An example}

\begin{figure}[htbp]
\begin{center}
\epsfig{file=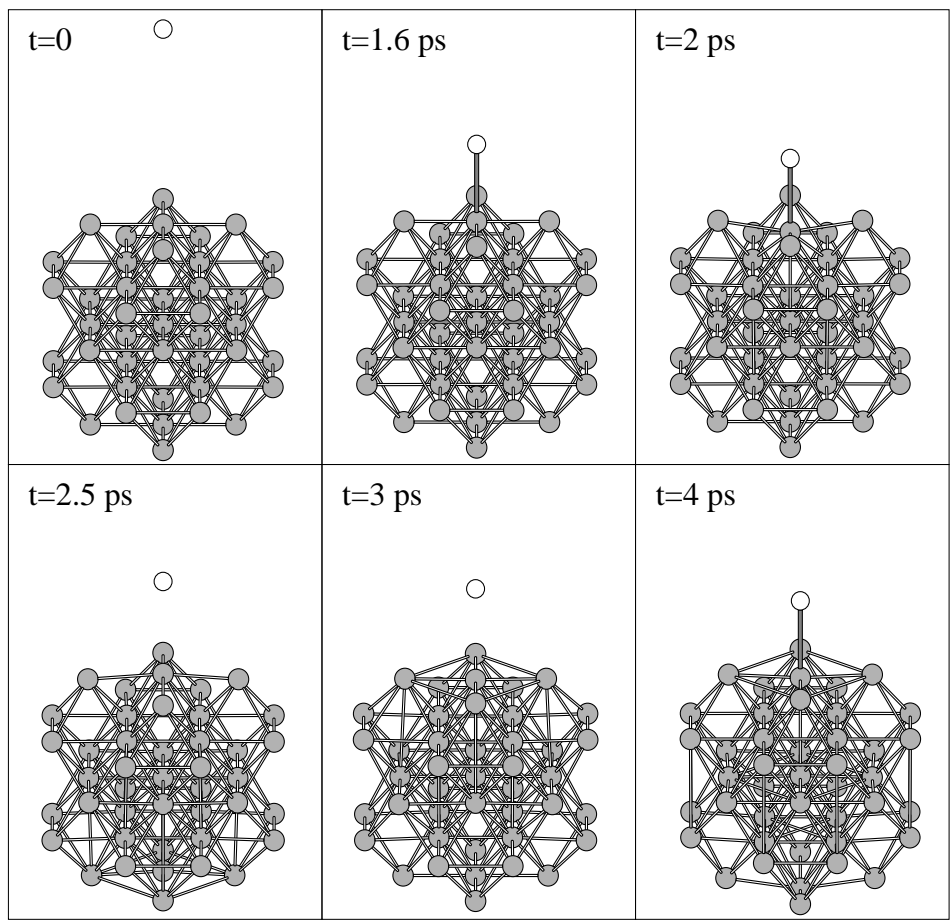,width=\linewidth}
\caption{\label{fig:NaAr43}
Example of dynamical deposition of a
Na atom on an Ar$_{43}$ cluster,  with initial 
kinetic energy E$_{0}$ = 13.6 meV. 
}
\end{center}
\end{figure}

We first consider, as an example, the case of the deposition of 
a Na atom on an Ar$_{43}$ cluster. This is illustrated in 
Figure~\ref{fig:NaAr43} where a few snapshots of the deposition 
process are presented.
In that test case, the Ar cluster presents to the
Na atom a rather small surface area of only 5 atoms
and the the Na atom is initially positioned above an Ar atom
(which shows strong core repulsion).
These two features differ as compared
to the case of Ar$_{86}$ illustrated in Figure \ref{fig:config} and
discussed later on in Section~\ref{sec:size}.
They should {\it a priori} make an attachment of the Na sodium more difficult
here. Still, the calculation 
shows that a faint binding takes place in this case within 
typically 3 ps. The Na atom loosely attaches to the Ar cluster
surface, since it keeps on bouncing over 7 ps with a decreasing
amplitude and around an average position of about 8 a$_0$, even a bit
closer than the NaAr dimer bond length of 9.5 a$_0$.
Mind that the initial Na kinetic energy $E_0$ was 13.6 meV,
in between the 5 meV binding energy of the NaAr dimer and the 50 meV
of the bonding in Ar bulk. Thus there is no surprise to observe the
creation of a transient NaAr bond and a very slight rearrangement of
the Ar cluster, which takes place over a time scale of order
of 5 ps. Actually, the Na atom is accelerated during its fall and 
gets, before the hit, a kinetic energy almost three times higher than
$E_0$ (this depends on its initial separation with the Ar first layer).
Then after the hit, the extra kinetic 
energy is absorbed by the Ar cluster both at the side 
of kinetic energy (about 15 meV) and in terms of potential energy in
its structure rearrangement. The residual
kinetic energy of Ar atoms as well as the potential energy consumed
in the Ar cluster rearrangement strongly depend on the available energy 
and the number of degrees of freedom. It is thus interesting to study the 
effect of variations of both these parameters.

\subsection{Dependence on kinetic energy}
\label{sec:Ekin}

\begin{figure}[htbp]
\begin{center}
\epsfig{file=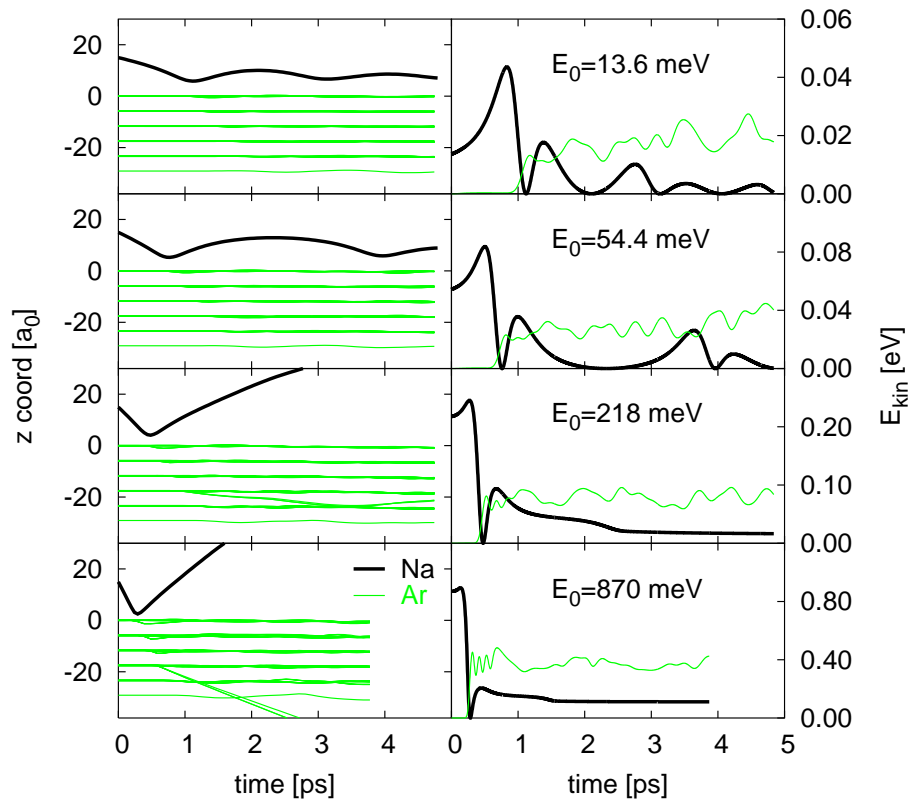,width=\linewidth}
\caption{\label{fig:NaAr86}
Dynamical deposition of Na (thick lines) on Ar$_{86}$ for four
different initial 
kinetic energies. The $z$-coordinates (left) and the
kinetic energies (right) are presented as a function of time.
}
\end{center}
\end{figure}

We first consider the influence of the initial kinetic energy 
of the Na projectile. The results are displayed in Figure 
 \ref{fig:NaAr86} for the deposition of a Na atom on Ar$_{86}$
at various energies (E$_0$ = 13.6 to 870 meV). 
It should first be noted that, whatever 
the initial kinetic energy no electronic 
emission is observed. We shall thus restrict the discussion to ionic 
and atomic  degrees of freedom. In the case of the 
two lowest impact  energies, one observes some bounces, and finally
a binding of the Na  to the surface 
at a typical distance of about $8-10$ $a_0$ from the first layer.
However for higher E$_{0}$, the Na is reflected by
the Ar cluster, and no attachment of the Na is observed, although the
Na is initially located above an interstitial position. 
This result can be confirmed by computing the Born-Oppenheimer 
surface of Na on an infinite Ar surface. The ground state surface 
exhibits a faint minimum around 7 $a_0$ above the Ar surface, 
which is fully compatible with our dynamical analysis 
on the finite Ar$_{86}$ cluster. Closer to the Ar surface, 
the Born-Oppenheimer surface exhibits a strong repulsion 
reflecting the core repulsion between Na and Ar. When the 
impact kinetic energy becomes too large the Na atom thus 
"misses" the faint minimum and directly explores the 
strongly repulsive part of the potential. The Na atom is then reflected 
by the cluster as observed in Figure \ref{fig:NaAr86}.
The energy threshold for neutral Na sticking seems to lie
between 0.05 and 0.2 eV. Note that in this range of kinetic energies,
the Ar cluster is not affected very much. This is particularly visible
on the time dependence of its kinetic energy, which exhibits a
rather smooth energy transfer from the Na to the Ar cluster. We also
observe the propagation of a soft shock wave in the substrate. This
participates to the slight heating  of the Ar cluster (to a few  tens of K), except
for the highest energy where some deep Ar atoms are emitted because of
the stronger wave propagating through the layers.

\subsection{Dependence on Ar cluster size}
\label{sec:size}

\begin{figure}[htbp]
\begin{center}
\epsfig{file=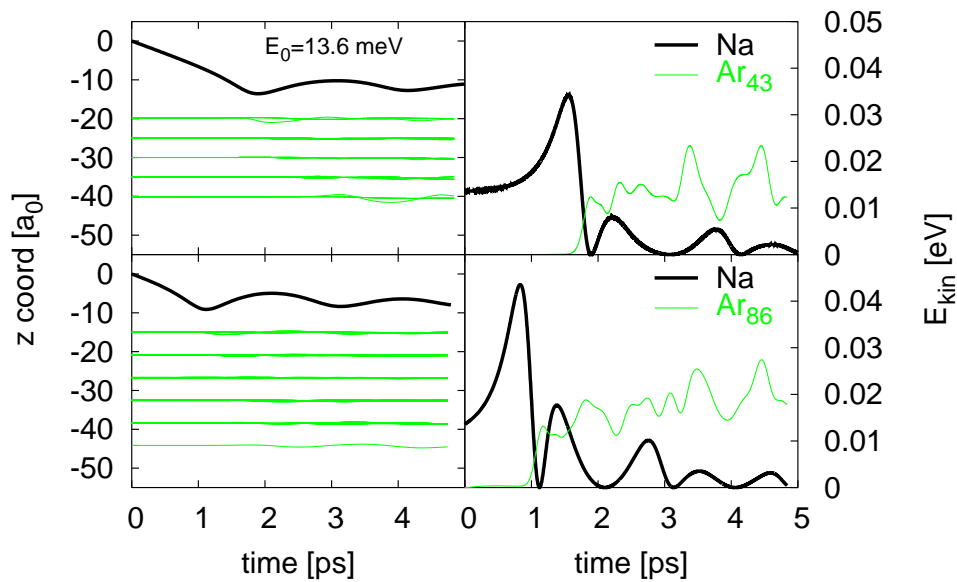,width=\linewidth}
\caption{\label{fig:matrix_size}
Deposit of a single Na atom (E$_0$ = 13.6) on Ar$_N$ systems of 
various sizes ($N =$ 43 and 86).  
Are presented as a function of time, the $z$ coordinates (left) and
the kinetic energies (right) of the Na atom (thick lines) and the Ar
cluster (thin curves).}
\end{center}
\end{figure}

We now consider the influence of the number of degrees of freedom 
on the capacity to dissipate the available energy. In practice, this amounts to 
test the influence of the Ar cluster size for constant initial kinetic
energy of the Na atom. The comparison is presented in 
Figure \ref{fig:matrix_size} where we plot Na and Ar positions 
and kinetic energies as a function of time during the 
deposition process of a Na atom (with initial kinetic energy E$_0$ of
13.6 meV) on Ar$_{43}$ and Ar$_{86}$. As already mentioned, two main
differences between both cases are to be noted. First 
the Na atom is initially above an Ar atom of the Ar$_{43}$, at a
distance of 20 $a_0$. In the case of Ar$_{86}$, the Na atom is above
an interstitial site and starts at a smaller distance from the Ar
first layer, namely 15 $a_0$. Due to their different sizes, the 
effective Ar surface exhibits only 5 atoms in the case of Ar$_{43}$ but
12 Ar atoms for  Ar$_{86}$.

Similar oscillating patterns are observed in both cases. 2 ps after
the impact, almost all kinetic energy 
of the Na atom is transfered to the Ar cluster, while the Ar kinetic
energy seems to reach an equilibrium and oscillates around a mean
value slightly higher than the initial kinetic energy $E_0$. 
Mind that the transferred energy is distributed equally over all Ar
atoms.  This means that, as expected, the larger the target cluster
size (thus the more available degrees of freedom), the smaller the
relative energy shared per Ar atom, and so the more moderate the
perturbation at their side.
Note, furthermore, that more Na kinetic energy is available at the time
of impact (maximum of Na kinetic energy in the right panels) for the
case of Ar$_{86}$. This means that the Na atom is accelerated faster
in that case as compared with Ar$_{43}$. The reason is that the larger
cluster and its larger surface area provide more attraction from
polarization potentials.  Nonetheless, in both cases, the Na atom
loosely binds to the surface, at about the same distance of about 8
$a_0$.

\section{Example of a finite cluster deposit}

\begin{figure}[htbp]
\begin{center}
\epsfig{file=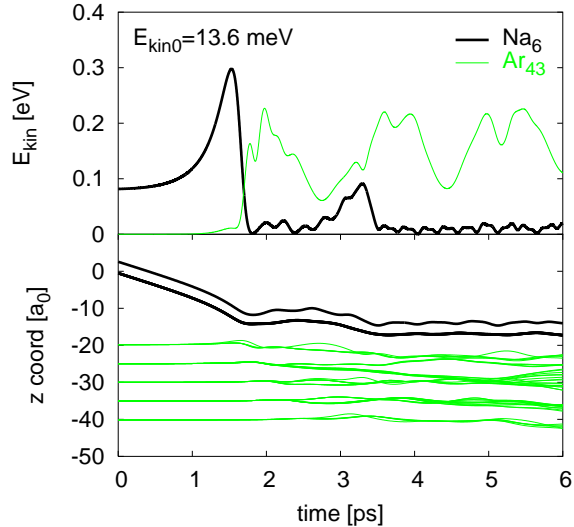,width=8cm}
\caption{\label{fig:Na6deposit}
Example of deposit of a finite Na$_6$ cluster on Ar$_{43}$.
}
\end{center}
\end{figure}

In order to complement the analysis in terms of Ar cluster size, it 
is also interesting to consider the case of the deposition 
of a full Na cluster, instead of a single atom. This again 
enhances the number of degrees of freedom. 
We shall consider comparable impact 
kinetic energy per Na atom. In Figure \ref{fig:Na6deposit} we 
consider an example of a deposition of 
a finite Na cluster (Na$_6$) on Ar$_{43}$. The initial 
kinetic energy of the cluster is $E_{\rm kin0}=13.6$ meV per Na atom. 
As in the case of a single Na atom (Figure \ref{fig:NaAr43}) 
the pinning process proceeds stepwise with a slight bounce
before the metal cluster finally attaches to the rare gas cluster.
At variance with the case of a single atom, though, one can furthermore  
analyze the evolution of the shape of the deposited Na cluster. The considered 
Na$_6$ cluster is primarily strongly oblate, consisting of a 
pentagon of 5 Na atoms topped by one central atom. 
One can see that during the deposition process, the cluster shape is little affected. 
This might have been expected in view of the ideal "flat" shape of 
the Na$_6$ cluster which already presents  a large contact area to the
Ar surface. But we  also found   
for other, and less favorable, geometries (e.g.
the nearly spherical 
Na$_8$ cluster) that the cluster shape remains basically intact during
deposition. The details of this scenario of 
course depend on the size of the Ar cluster target and on the 
initial kinetic energy of the impinging Na cluster, but 
qualitatively the example displayed in Figure \ref{fig:Na6deposit}
turns out to be quite typical. These results 
 will be presented elsewhere \cite{wet}. 

\section{Charge effects}

\begin{figure}[htbp]
\begin{center}
\epsfig{file=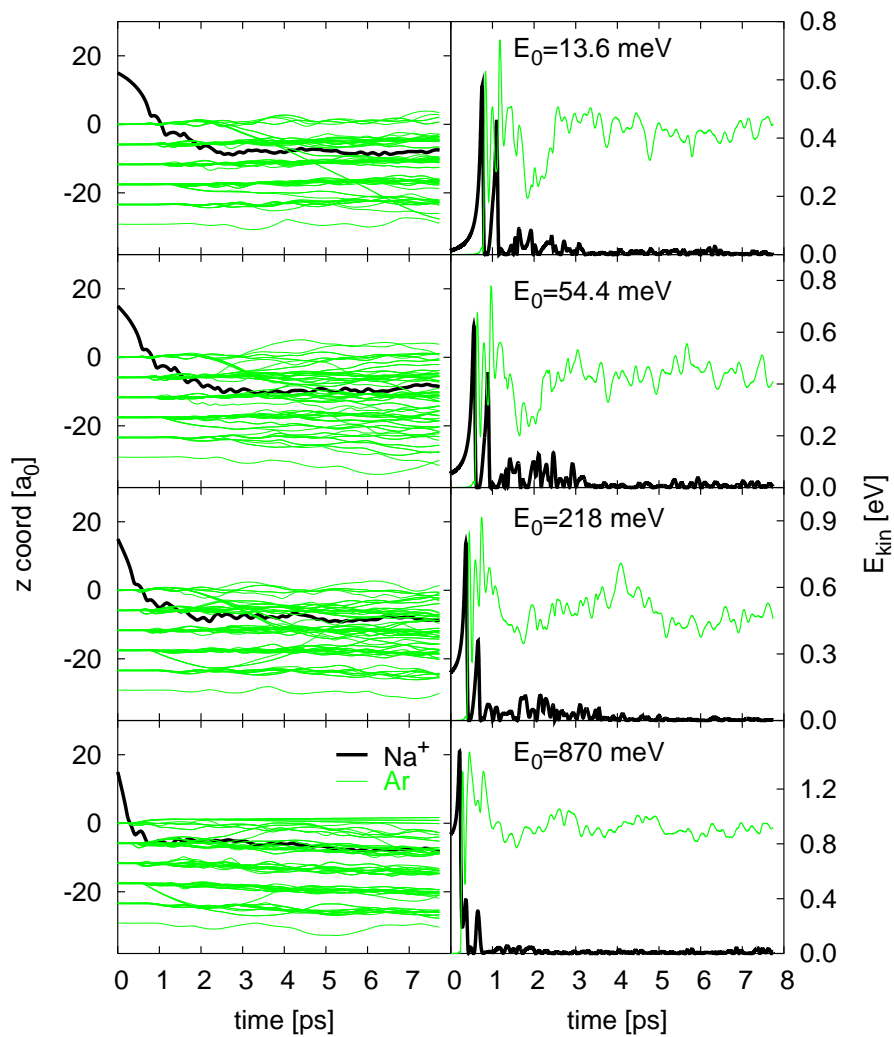,width=\linewidth}
\caption{\label{fig:NapAr86}
Same as in Figure~\ref{fig:NaAr86}, in the case of Na$^+$ on
Ar$_{86}$. 
}
\end{center}
\end{figure}

As a final point, we want now to analyze the influence of charge on 
the deposition process. We come back to the simple case of a
single Na atom and consider the deposition of the corresponding 
Na$^+$ ion on Ar$_{86}$, at various initial kinetic energies. 
Born Oppenheimer calculations of Na$^+$ in contact to an Ar surface 
show that the attachment is much stronger (and closer to the surface) 
than in the case of the neutral species. This can be easily 
understood by remembering the key  role played by the attractive
Ar polarization potentials. The finite charge of the Na$^+$ ion strongly 
polarizes the surface and thus enhances the binding as compared to the 
neutral case. Of course short range repulsion remains present but 
will take the lead only on shorter distances. The attachment is thus 
expected to be stronger and closer to the surface.

Figure \ref{fig:NapAr86} confirms the expectation.
The Na$^+$ ion is practically swallowed by the Ar cluster 
and the Ar cluster itself undergoes stronger rearrangements.
For the smallest $E_0$ presented in the top
panels of Figure~\ref{fig:NapAr86}, two of the six surface 
atoms are finally ejected
from the Ar cluster (they are visible by the light lines going
straight through Ar layers in the $z$-coordinate panel and having
reached
the lower end at 7 ps). The four other
Ar atoms of the (strongly perturbed) surface remain bound to the whole cluster
and participate to its strong rearrangement. In the case of intermediate
$E_0$, the six Ar atoms are lifted from the first layer but still
stick to the edges of the Ar cluster. The fact that the Na$^+$ goes
deeper for $E_0=54.4$ meV seems to be due to the final ejection of an
Ar atom. In the latter three cases, these six atoms get about 10\% of
the total Ar kinetic energy. Finally, for the highest $E_0$,
the first layer just explodes after the impact. The  six outmost Ar
atoms absorb up to 50\% of the vertical shock in term of kinetic
energy transfer, and then follow a radial motion in an horizontal
plane (see the thin horizontal lines around $z=0$ in the bottom left
panel of Fig.\ref{fig:NapAr86}). 
In all cases, the removal of the six
atoms from the first Ar layer, either outside or at the edges of the
Ar cluster, allows some Ar atoms deeper inside the substrate to move
upwards, thus leaving a large vacancy, so that  the Na$^+$ can
penetrate even under the second layer. These results suggest that 
the number of
layers under the Ar surface probably does not play an important role
in the Na$^+$ inclusion. 
More important is the surface's mobility for rearrangements.

\section{Conclusion}

In this paper, we have presented results on the deposition of
a Na atom, a Na$^+$ ion, and a Na$_6$ cluster
on a dynamically polarizable Ar substrate represented by 
finite Ar clusters of various  sizes. We have used 
time-dependent
density-functional theory for the Na electrons 
coupled to molecular dynamics for the treatment 
of Na ions and Ar atoms.
We have presented systematic results as a function of 
Ar cluster size and kinetic energy of the impinging Na atom. 
We have found that the neutral Na is not 
likely to penetrate into the Ar matrix and sticks to the Ar surface for
initial kinetic energy lower than $\sim 0.2$ meV while being
reflected for larger impact kinetic energies. 
In case of the
positively charged Na$^+$, inclusion is observed, whatever the initial
kinetic energy. The Ar matrix (or the finite Ar cluster) then
undergoes strong perturbations and ejects one or more atoms to create
a vacancy for the Na$^+$ inclusion. 
As a first exploratory example, we have also studied
the deposition of a neutral Na$_6$ cluster. As for the neutral
atom, we see also
a binding to the surface and no penetration into  the Ar substrate. 
Somewhat surprisingly, there is only little perturbation  of  the
Na cluster internal structure.
Continued systematic investigations on metal cluster
deposition in Ar substrate are in progress.

\bigskip

Acknowledgments: This work was supported by the DFG, project nr. RE
322/10-1, the French-German exchange program PROCOPE nr. 07523TE, the
CNRS Programme ``Mat\'eriaux'' (CPR-ISMIR), Institut Universitaire de
France, the Huomboldt foundation and a Gay-Lussac price, and has
benefited from the CALMIP (CALcul en MIdi-Pyr\'en\'ees) computational
facilities. 


\appendix{The Na-Ar energy functional in detail}
\label{sec:enfundetail}


%
The degrees of freedom of the model are:\\
%
\begin{tabular}{ll}
$\{\varphi_n({\bf r}),n=1...N_{\rm el}\}$&
wavefunctions of cluster electrons\\
%
$\{{\bf R}_I,I=1...N_{\rm ion}\}$ & coordinates
of cluster's  Na$^+$ ions\\
%
$\{{\bf R}_a,a=1...N_{\rm Ar}\}$& 
coordinates of Ar atoms (cores Ar$^{Q+}$)\\
%
$\{{\bf R'}_a,a=1...N_{\rm Ar}\}$& 
coordinates of the Ar valence clouds\\
\end{tabular}

An Ar atom is described by two constituents with opposite
charge, positive Ar core and negative Ar valence cloud, which allows 
a correct description of polarization dynamics. 
We associate a Gaussian charge charge distribution to
both constituents having a width of the order of the 3p shell in Ar,
in the spirit of \cite{Dup96}. 
The dynamical polarizability of the Na$^+$ ions is neglected
and we treat them simply as charged point particles.


The total energy of the system is composed as:
$$
  E_{\rm total}
  =
  E_{\rm Na cluster}
  +
  E_{\rm Ar}
  +
  E_{\rm coupl}
  +
  E_{\rm VdW}
  \quad,
$$
%
The energy of the Na cluster $E_{\rm Na cluster}$ consists out of
TDLDA (with SIC) for the electrons, MD for ions, and a coupling of
both by soft, local pseudo-potentials, see \cite{Cal00,Rei03a} for details.
The Ar system and its coupling to the clusters is described by
\begin{eqnarray*}
\label{eq:Na-cluster}\\
  E_{\rm Ar}
  &=&
  \sum_a \frac{{\bf P}_a^2}{2M_{\rm Ar}} 
  +
  \sum_a \frac{{{\bf P}'_{a}}^2}{2m_{\rm Ar}}
  +
  \frac{1}{2} k_{\rm Ar}\left({\bf R}'_{a}-{\bf R}_{a}\right)^2
\nonumber\\
  &&  
  +
  \sum_{a<a'}
  \left[
    \int d{\bf r}\rho_{{\rm Ar},a}({\bf r})
    V^{\rm(pol)}_{{\rm Ar},a'}({\bf r})
    +
    V^{\rm(core)}_{\rm ArAr}({\bf R}_a - {\bf R}_{a'})
  \right]
  \quad,
\\
  E_{\rm coupl}
  &=&
  \sum_{I,a}\left[
    V^{\rm(pol)}_{{\rm Ar},a}({\bf R}_{I})
    +
    V'_{\rm NaAr}({\bf R}_I - {\bf R}_a)
  \right]
\nonumber\\
  &&
  +
  \int d{\bf r}\rho_{\rm el}({\bf r})\sum_a \left[
    V^{\rm(pol)}_{{\rm Ar},a}({\bf r})
    +
    W_{\rm elAr}(|{\bf r}-{\bf R}_a|)
  \right]
  \quad,
\\
  V^{\rm(pol)}_{{\rm Ar},a}({\bf r})
  &=&
  e^2{q_{\rm Ar}^{\mbox{}}}
  \Big[
   \frac{\mbox{erf}\left(|{\bf r}\!-\!{\bf R}^{\mbox{}}_a|
          /\sigma_{\rm Ar}^{\mbox{}}\right)}
        {|{\bf r}\!-\!{\bf R}^{\mbox{}}_a|}
   -
   \frac{\mbox{erf}\left(|{\bf r}\!-\!{\bf R}'_a|/\sigma_{\rm Ar}^{\mbox{}}\right)}
        {|{\bf r}\!-\!{\bf R}'_a|}
  \Big]
  \quad,
\label{eq:Arpolpot}
\\
  W_{\rm elAr}(r)
  &=&
  e^2\frac{A_{\rm el}}{1+e^{\beta_{\rm el}(r - r_{\rm el})}}
\label{eq:VArel}\\
  V_{\rm ArAr}^{\rm (core)}(R)
  &=& 
  e^2 A_{\rm Ar}\Bigg[
  \left( \frac{R_{\rm Ar}}{R}\right)^{12}
 -\left( \frac{R_{\rm Ar}}{R}\right)^{6}
  \!\Bigg]
\label{eq:VArAr}
\\
  V'_{\rm ArNa}(R)
  &=&
  e^2\Bigg[
  A_{\rm Na} \frac{e^{-\beta_{\rm Na} R}}{R}
  -
  \frac{2}{1+e^{\alpha_{\rm Na}/R}}
  \left(\frac{C_{\rm Na,6}}{R^6} + \frac{C_{\rm Na,8}}{R^8}\right)
  \Bigg]
\label{eq:VpArNa}
\\
  &&
  4\pi \rho_{{\rm Ar},a}
  =
  \Delta V^{\rm(pol)}_{{\rm Ar},a}
\\
  E_{\rm VdW}
  &=&  
  e^2\frac{1}{2} \sum_a \alpha_a
  \Big[
    \frac{
       \left(\int{d{\bf r} {\bf f}_a({\bf r}) \rho_{\rm el}({\bf r})}\right)^2
         }{N_{\rm el}}
      - \int{d{\bf r} {\bf f}_a({\bf r})^2 \rho_{\rm el}({\bf r})}
  \Big]
  \;,
\label{eq:EvdW}
\\
  &&
  {\bf f}_a({\bf r})
  =
  \nabla\frac{\mbox{erf}\left(|{\bf r}\!-\!{\bf R}^{\mbox{}}_a|
          /\sigma_{\rm Ar}^{\mbox{}}\right)}
        {|{\bf r}\!-\!{\bf R}^{\mbox{}}_a|}
  \quad.
\label{eq:effdip}
\\
  &&
  \mbox{erf}(r)
  = 
  \frac{2}{\sqrt{\pi}}\int_0^r dx\,e^{-x^2}
  \quad.
\end{eqnarray*}
The various contributions are calibrated from independent
sources, with a final fine tuning to the NaAr dimer
modifying only the term $W_{\rm elAr}$. The parameters are
summarized in  the table. The third column of the table
indicates the source for the parameters. 

\begin{table}
\begin{tabular}{|l|l|l|}
\hline
 \rule[-8pt]{0pt}{22pt}
 $V^{\rm(pol)}_{{\rm Ar},a}$
&
 $q_{\rm Ar}
  =
  \frac{\alpha_{\rm Ar}m_{\rm el}\omega_0^2}{e^2}$
 \;,\;
 $k_{\rm Ar}
  =
  \frac{e^2q_{\rm Ar}^2}{\alpha_{\rm Ar}}$
 \;,\;
 $m_{\rm Ar}=q_{\rm Ar}m_{\rm el}$
&
 $\alpha_{\rm Ar}$=11.08$\,{\rm a}_0^3$
\\
 \rule[-12pt]{0pt}{22pt}
 &
 $\sigma_{\rm RG}
  =
  \left(\alpha_{\rm Ar}\frac{4\pi}{3(2\pi)^{3/2}}  \right)^{1/3}$
&
 \raisebox{12pt}{$\omega_0 = 1.755\,{\rm Ry}$}
\\
\hline
 \rule[-6pt]{0pt}{18pt}
 $W_{\rm elAr}$
&
 $A_{\rm el}$=0.47  
 \;,\;
 $\beta_{\rm el}$=1.6941\,/a$_0$  
 \;,\;
 $r_{\rm el}=$2.2 a$_0$ 
&
 fit to NaAr
\\ 
\hline
 \rule[-8pt]{0pt}{22pt}
$V^{\rm(core)}_{\rm ArAr}$
&
 $A_{\rm Ar}$=$1.367*10^{-3}$ Ry
 \;,\;
 $R_{\rm Ar}$=6.501 a$_0$ 
&
fit to bulk Ar
\\
\hline
 \rule[-6pt]{0pt}{18pt}
 $V'_{\rm ArNa}$
&
 $\beta_{\rm Na}$= 1.7624 a$_0^{-1}$
 \;,\;
 $\alpha_{\rm Na}$= 1.815 a$_0$
 \;,\;
 $A_{\rm Na}$= 334.85 
&
\\
 \rule[-6pt]{0pt}{18pt}
&
 $C_{\rm{Na},6}$= 52.5 a$_0^6$
 \;,\;
 $C_{\rm Na,8}$= 1383 a$_0^8$
&
after \cite{Rez95}
\\
\hline
\end{tabular}
\caption{
Parameters for the various model potentials.
}
\label{tab:params}
\end{table}

The (most important) polarization potentials are described by 
a valence electron cloud oscillating against the raregas core
ion. Its parameters are: 
$q_{\rm Ar}$ the effective charge of valence cloud, 
$m_{\rm Ar}=q_{\rm Ar}m_{\rm el}$ the effective mass of valence cloud, 
$k_{\rm Ar}$ the restoring force for dipoles, and 
$\sigma_{\rm Ar}$ the width of the core and valence clouds.
The $q_{\rm Ar}$ and $k_{\rm Ar}$ are adjusted to reproduce the dynamical
polarizability $\alpha_D(\omega)$ of the Ar atom at low frequencies,
namely the static limit 
$\alpha_D(\omega\!=\!0)$ and the second derivative
of $\alpha''_D(\omega\!''=\!0)$.
The width $\sigma_{\rm Ar}$ is determined consistently such that the
restoring force from the folded Coulomb force (for small
displacements) reproduces the spring constant $k_{\rm Ar}$.

The short range repulsion is provided by the various core potentials.
For the Ar-Ar core interaction we employ a Lennard-Jones type
potential with parameters reproducing binding properties of bulk Ar. 
The Na-Ar core potential is chosen according to
\cite{Rez95}, within properly avoiding double counting of the 
 the dipole polarization-potential.

The pseudo-potential $W_{\rm elAr}$ for the electron-Ar core repulsion
has been modeled according to the proposal of \cite{Dup96} with
a final slight adjustment to the Na-Ar
dimer,  data taken from from \cite{Gro98} and \cite{Rho02a}.

The Van-der-Waals energy $E_{\rm VdW}$ is a correlation from the
dipole excitation in the Ar atom coupled with a dipole excitation in
the cluster. We exploit that $\omega_{\rm Mie}\ll\Delta E_{\rm Ar}$
which simplifies the term to the variance of the dipole operator in
the cluster, using again the regularized dipole operator ${\bf f}_a$
corresponding to the smoothened Ar charge distributions. The full
dipole variance is simplified in terms of the local variance.

\end{document}